# Nanoparticulate Metal Oxide Top Electrode Interface Modification Improves the Thermal Stability of Inverted Perovskite Photovoltaics

**Ioannis T. Papadas [1], Fedros Galatopoulos [1], Gerasimos S. Armatas [2], Nir Tessler [3] and Stelios A. Choulis [1],***

[1] Molecular Electronics and Photonics Research Unit, Department of Mechanical Engineering and Materials Science and Engineering, Cyprus University of Technology, Limassol 3603, Cyprus

[2] Department of Materials Science and Technology, University of Crete, 71003 Heraklion, Greece

[3] Sara and Moshe Zisapel Nano-Electronic Center, Department of Electrical Engineering, Technion-Israel Institute of Technology, Haifa 32000, Israel

* Correspondence: stelios.choulis@cut.ac.cy

**Abstract:** Solution processed $\gamma$-Fe$_2$O$_3$ nanoparticles via the solvothermal colloidal synthesis in conjunction with ligand-exchange method are used for interface modification of the top electrode in inverted perovskite solar cells. In comparison to more conventional top electrodes such as PC(70)BM/Al and PC(70)BM/AZO/Al, we show that incorporation of a $\gamma$-Fe$_2$O$_3$ provides an alternative solution processed top electrode (PC(70)BM/$\gamma$-Fe$_2$O$_3$/Al) that not only results in comparable power conversion efficiencies but also improved thermal stability of inverted perovskite photovoltaics. The origin of improved stability of inverted perovskite solar cells incorporating PC(70)BM/ $\gamma$-Fe$_2$O$_3$/Al under accelerated heat lifetime conditions is attributed to the acidic surface nature of $\gamma$-Fe$_2$O$_3$ and reduced charge trapped density within PC(70)BM/ $\gamma$-Fe$_2$O$_3$/Al top electrode interfaces.

## 1. Introduction

Over the last couple of years, perovskite solar cells proved to be a strong candidate for photovoltaic (PV) technologies while showing a considerable improvement in power conversion efficiency (PCE) from 3.8% to 22.7% [1,2]. The attractiveness of perovskite solar cells as a novel research field is a result of several features such as high carrier mobility [3], long carrier diffusion length [4], high absorption coefficient [5], tunable bandgap [6,7], and low exciton binding energy [8,9]. These characteristics allow perovskite solar cells to achieve high PCE while being compatible with solution processed fabrication techniques at low temperatures [10]. When considering the viability of any novel PV technology for competing in the silicon dominated industry, achieving high PCE is not enough. Ensuring long term stability is also equally important. Even though perovskite solar cells have achieved promising PCEs, poor stability often hinders this PV technology to reach its full potential. Perovskite-based solar cells still struggle to achieve good lifetime performance under moisture, heat and light soaking conditions. For instance, the perovskite crystal can decompose in the presence of water molecules into $PbI_2$ through mobilization of the organic species [11–13]. In order to improve the humidity stability of PSCs several methods were reported, such as the usage of hydrophobic thiols [11]. In one of the most recent reports, Fan et. al. utilized a cerium oxide bilayer ($CeO_x$) to achieve a 200-h lifetime under ambient conditions [14]. The perovskite crystal also shows poor intrinsic stability in the presence of $O_2$ under illumination. In such a case, superoxide ions ($O_2^-$) can be formed upon irradiation of $CH_3NH_3X_3$ (X = Cl, Br, I halides) which causes the decomposition of the perovskite crystal structure via the deprotonation of methylammonium cation ($MA^+$) [13, 15]. Improvements in the light stability of perovskite-based solar cells have been reported from several research groups by doping the [6,6]-phenyl-butyric acid methyl ester (PCBM) with graphene quantum dots (GQDs) [16] or by adopting devices based on mixed cations, such as formamidinium-cesium [$FA_{0.83}Cs_{0.17}Pb(I_{0.8}Br_{0.2})_3$] [17]. As previously mentioned, another important parameter that can affect the stability of perovskite-based solar cells is heat. The perovskite crystal was reported to decompose when exposed to oxygen atmospheres in conjunction with an elevated temperature of 85 °C [18]. To this end, several approaches have been suggested in order to improve the heat stability of perovskite-based solar cells. These include using thermally stable nickel oxide (NiO) and PCBM as the hole transporting layer (HTL) and electron transporting layer (ETL), respectively [19], or even using a thin Cr layer between the HTL and active layer to block the diffusion of Au into the active layer, which was shown to take place even at temperatures as low as 70 °C [20].

Amongst the different buffer layers that have been used, both for improving the PCE and stability of PSCs, metal oxides have been a relatively popular choice [21]. Al-doped zinc oxide (AZO) has been shown to act as an efficient ETL for perovskite-based solar cells [6], whereas incorporations of tin oxide ($SnO_x$) on top of AZO has also been reported proven to improve both humidity and heat stability of p-i-n structure perovskite-based devices [22]. Recent publications have also incorporated delafossites, such as $CuGaO_2$ [23], and spinel structure oxides, such as $NiCo_2O_4$ [24], as efficient HTLs.

Iron(III) oxide ($Fe_2O_3$) has recently become a very attractive material for environmental applications, due to its high chemical stability, low cost, abundance, non-toxicity and magnetic separation ability [25,26]. Examples include photocatalysis, adsorption, reduction, water splitting and organic synthesis area [27–34]. Recently, $Fe_2O_3$ is explored as a potential photoanode material in visible-light-driven water splitting due to its visible light response (bandgap energy ~2–2.2 eV) and appropriate valence band ($E_v$) position (~2.5 V vs. NHE at pH = 1) [35]. There are continuous efforts towards the synthesis of well-defined $Fe_2O_3$ nanostructures. A large variety of $Fe_2O_3$ nanoparticles (NPs) with fine control over the size (from 2 to 100 nm) and shape (from cubic, to spherical and to nanosheets) have been successfully implemented [36]. Unfortunately, the implementation of such nanomaterials, especially in planar perovskite solar cells, is hobbled by their strong tendency to form large agglomerates in solution. In addition, the morphology of the obtained segregated assemblies is irregular, and thus new strategies to access stable monodispersed solutions of NPs and their usage in solution processed techniques for the fabrication of low-cost hybrid perovskite solar cells are indispensable. $Fe_2O_3$ has been previously used in normal n-i-p structured perovskite solar cells for replacing the more conventional titanium dioxide ($TiO_2$), showing improved PCE [37]. A similar approach utilized $Fe_2O_3$ nano islands to enhance the photostability of n-i-p perovskite solar cells by replacing the $TiO_2$ [38]. The synergistic effect of fullerene/$Fe_2O_3$ was also explored as ETL, which resulted in improved PCE and ambient stability of devices by improving the crystallinity of the perovskite layer [39]. Although $Fe_2O_3$ has been a very common and popular metal oxide, it is relatively unexplored as an interfacial layer in the field of inverted perovskite solar cells.

In this work, we report improved thermal stability for inverted (p-i-n) perovskite-based solar cells using colloidal synthesized nanoparticulate $\gamma$-$Fe_2O_3$ interfacial modification. By incorporating a thin (~15 ± 5 nm) $\gamma$-$Fe_2O_3$ interfacial layer between PC(70)BM and Al we report improved lifetime performance under accelerated heat conditions (60 °C) and $N_2$ atmosphere for inverted perovskite-based solar cells. Using frequency-dependent impedance spectroscopy we show that the top electrode $\gamma$-$Fe_2O_3$ based interface modification improved heat stability due to the reduced charge trap density, as evident by the insensitivity of Vbi to frequency changes.Subsequently, a more intimate interface is formed between the $\gamma$-$Fe_2O_3$ ETL and the top metal electrode compared to inverted perovskite solar cells using more conventional ETLs such as plain PC(70)BM and aluminum-doped zinc oxide (AZO) under identical (accelerated heat lifetime) conditions. In contract to AZO, the surface properties of $\gamma$-$Fe_2O_3$ appear to be beneficial for the protection of perovskite active layer from the deprotonation of the methylamine. The experimental results presented highlight the potential of the proposed $\gamma$-$Fe_2O_3$ as an effective interface modification for inverted perovskite-based solar cells.

**2. Experimental Section**

*2.1. Materials*

*Chemical reagents:* Iron (III) chloride hexahydrate ($FeCl_3.6H_2O$, ≥97%), Oleylamine (85%), Sodium oleate (80%), Oleyl alcohol, (85%), diphenyl ether (99%), *N*,*N*-dimethylformamide (DMF, 99.9%), acetone, acetonitrile (99.9%), hexane (95%), toluene (99.7%) and absolute ethanol (98%) were purchased from Sigma-Aldrich Chemical (Darmstadt, Germany) Co. Nitrosonium tetrafluoroborate ($NOBF_4$, 97%) was purchased from Acros Organics (San Mateo, CA, USA). Ultrapure water was produced by a milli-Q Academic system, Millipore (Burlington, Massachusetts, United States). All solutions were prepared with analytical grade chemicals and ultrapure milli-Q water with a conductivity of 18.2 μS cm$^{−1}$.

*2.2. Synthesis of γ-Fe₂O₃ NPs*

Iron oxide NPs with an average diameter of ~5 nm was prepared according to the previously reported procedure [28], with slight modifications highlighted below.

Briefly, FeCl$_3$.6H$_2$O (5.4 g, 20 mmol) and sodium oleate (18.25 g, 60 mmol) were added to a mixture of ethanol (40 mL), deionized water (30 mL), and hexane (70 mL). The mixture was refluxed at 70 °C for 4 h, then the upper brown hexane solution containing iron–oleate complex was separated, washed three times with deionized water (50 mL), and dried under vacuum at 60 °C for 12 h, yielding a dark brown, oily iron oleate complex. Finally, the iron oleate complex (10.8 g) was dissolved in 60 g of diphenyl ether with the addition of oleyl alcohol (19.4 g). Under N$_2$ flux, this mixture under stirring was heated to 105 °C at a ramp rate of 2 °C min$^{-1}$ and kept for 10 min at this temperature to eliminate H$_2$O and adsorbed O$_2$ [28]. For the work reported in this paper the mixture was heated to 220 °C at a ramp rate of 2 °C min$^{-1}$ and the reaction could proceed for 20 min at this temperature. The resulting black brown nanocrystal solution was left to cool at room temperature. Then, the γ-Fe$_2$O$_3$ NPs were precipitated with 50 mL acetone followed by centrifugation at 14.500 rpm for 20 min and re-dispersed in hexane to form a stable colloidal solution (10 mg mL$^{-1}$).

*2.3. Preparation of Ligand-Stripped γ-Fe$_2$O$_3$ NPs*

The surface of γ-Fe$_2$O$_3$ NPs was modified with NOBF$_4$ using a ligand-exchange reaction. Briefly, equal volumes (5 mL) of colloidal γ-Fe$_2$O$_3$ NPs in hexane and 0.01 M NOBF$_4$ in acetonitrile were mixed and the resulting mixture was kept under stirring at room temperature until the NPs were transferred to the acetonitrile phase (typically within ~1 h). The BF$_4^-$ capped Fe$_2$O$_3$ NPs were then collected by precipitation with toluene followed by centrifugation, dried under vacuum at 40 °C for 12 h, and dispersed in ethanol to form a stable colloidal solution (5 mg mL$^{-1}$).

*2.4. Device Fabrication*

The inverted (p-i-n) solar cells under study had the structures ITO/PEDOT:PSS/CH$_3$NH$_3$PbI$_3$/PC$_{70}$BM/Al, ITO/PEDOT:PSS/CH$_3$NH$_3$PbI$_3$/PC$_{70}$BM/AZO/Al and ITO/PEDOT:PSS/CH$_3$NH$_3$PbI$_3$/PC$_{70}$BM/γ-Fe$_2$O$_3$/Al, respectively. ITO-patterned glass substrates (sheet resistance = 4 Ω sq$^{-1}$) were cleaned using an ultrasonic bath for 10 min in acetone followed by 10 min in isopropanol. The PEDOT:PSS films were prepared using the Al4083 solution from Heraeus. PEDOT:PSS was filtered using 0.22 μm polyvinylidene difluoride (PVDF) filters before coating. The PEDOT:PSS films were coated using spin-coating at 6000 rpm for 30 s followed by annealing at 150 °C for 10 min. The perovskite solution was prepared using a mixture (1:1) of PbI$_2$ from Alfa Aesar and MAI from GreatCell Solar. The mixture was dissolved in a solvent (7:3) of γ-butyrolactone and dimethyl sulfoxide (DMSO). The solution was stirred at 60 °C for 1 h. The perovskite solution was left to cool at room temperature inside the glove box followed by filtering using a 0.22 μm PVDF filter. The perovskite films were coated using a three-step spin-coating process: first step 500 rpm for 5 s, second step 1000 rpm for 45 s, and third step 5000 rpm for 45 s. During the third step, after the first 20 s of the duration of the step, 0.5 mL toluene was dropped onto the spinning substrate to achieve the rapid crystallization of the films. The resulting perovskite films were annealed at 100 °C for 10 min. The PC(70)BM film was then coated using a solution of 20 mg mL$^{-1}$ in chlorobenzene to achieve fullerene thickness of 70 nm, at 1000 rpm for 30 s. The AZO solution used was a commercial solution (N-20X) from Avantama Ltd. AZO films were coated using spin-coating at 3.000 rpm for 30 s. To form a 15 ± 5 nm electron transporting layer (ETL) of γ-Fe$_2$O$_3$, the γ-Fe$_2$O$_3$ NPs dispersion (5 mg L$^{-1}$ in ethanol) used in this study was dynamically spin-coated at 3000 rpm for 30 s on top of the PC(70)BM films. The thicknesses of the device layers were accurately measured with profilometer. The Al metal was finally thermally evaporated achieving a thickness of 100 nm. The devices were encapsulated directly after the evaporation of the metal contact, in the glove box using a glass coverslip and an encapsulation epoxy resin (Ossila E131) activated by 365 nm UV-irradiation.

*2.5. Characterization*

The thicknesses and the surface profile of the device layers were measured with a Veeco Dektak 150 profilometer (Hong Kong, China). Wide-angle XRD patterns were collected on a PANanalytical X'pert Pro MPD X-ray diffractometer (Malvern, UK) equipped with a Cu (λ = 1.5418 Å) rotating anode operated at 40 mA and 45 kV, in the Bragg-Brentano geometry. The current density-voltage (J/V) characteristics were characterized by a Botest LIV Functionality Test System (Kreuzwertheim, Germany). Both forward (short circuit open circuit) and reverse (open circuit short circuit) scans were measured with 10 mV voltage steps and 40 ms of delay time. For illumination, a calibrated Newport Solar simulator (Irvine, California, USA) equipped with a Xe lamp was used, providing an AM1.5G spectrum at 100 mWcm$^{-2}$ as measured by a certified oriel 91,150 V calibration cell. A custom-made shadow mask was attached to each device prior to measurements to accurately define the corresponding device area. EQE measurements were performed by Newport System (Irvine, California, USA), Model 70356_70316NS. Impedance spectroscopy was performed using a Metrohm Autolab PGSTAT 302N equipped with FRA32M module (Herisau, Switzerland). To extract the Nyquist plots, the devices were illuminated using a red LED at 625 nm and 100 mw/cm$^2$. A small AC perturbation voltage of 10 mV was applied, and the current output was measured using a frequency range of 1 MHz to 1 Hz. The steady-state DC bias was kept at 0 V. Mott-Schottky measurements on γ-Fe$_2$O$_3$ films were performed in a 0.5 M Na$_2$SO$_4$ aqueous electrolyte (pH = 7) using a Metrohm Autolab PGSTAT 302N potentiostat. A three-electrode set-up, with a platinum plate (1.0 × 2.0 cm$^2$) and a silver-silver chloride (Ag/AgCl, 3M KCl) as the counter and reference electrodes, respectively, was adopted to study the samples. The capacitance of the semiconductor/electrolyte interface was obtained at 1 kHz, with 10 mV AC voltage perturbation. All the experiments were conducted under dark conditions. The measured potential vs the Ag/AgCl reference electrode was converted to the normal hydrogen electrode (NHE) scale using the formula: $E_{NHE} = E_{Ag/AgCl} + 0.210$ V. The working electrode for impedance-potential measurement was fabricated as follows: ~10 mg of γ-Fe$_2$O$_3$ NPs was dispersed in 1 mL DI water and the mixture was subjected to sonication in a water bath until a uniform suspension was formed. After that, 100 µL of the suspension was drop-casted onto the surface of fluorine-doped tin oxide (FTO, 9 Ω/sq) substrate, which was masked with an epoxy resin to expose an effective area of 1.0 × 1.0 cm$^2$. The sample was dried in a 60 °C oven for 30 min. Photoluminescence (PL) spectrum was obtained at room temperature on a Jobin-Yvon Horiba FluoroMax-P (SPEX) spectrofluorimeter (Singapore) equipped with a 150 W Xenon lamp and operated from 300 to 900 nm The C–V measurements for the Mott–Schottky plots for full devices were performed under dark using a voltage range of −1 to 1 V at a constant frequency of f = 5 kHz. Transmittance and absorption measurements were performed with a Shimadzu UV-2700 UV–vis spectrophotometer (Kyoto, Japan). Diffuse reflectance UV–vis spectra were recorded at room temperature with a Shimadzu UV-2700 UV-Vis optical spectrophotometer, using powder $BaSO_4$ as a 100% reflectance standard. Reflectance data were converted to absorption (α/S) data according to the Kubelka-Munk equation: α/S = (1 − R)$^2$/(2R) [40], where R is the reflectance and α, S are the absorption and scattering coefficients, respectively. Atomic force microscopy (AFM) images were obtained using a Nanosurf easy scan 2 controllers in tapping mode. Transmission electron microscopy (TEM) images were taken with a JEOL Model JEM-2100 electron microscope operating at 200 kV accelerated voltage. Samples for TEM analysis were prepared by drying an ethanolic dispersion of the particles on a holey carbon-coated Cu grid.

### 3. Results and Discussion

*3.1. γ-Fe$_2$O$_3$ Interface Modification for Inverted Perovskite Solar Cells*

In this work, we have used solvothermal colloidal synthesis in conjunction with the ligand-exchange method to isolate uniform γ-Fe$_2$O$_3$ NPs, well-dispersed in polar solvents. In particular, we utilized nitrosonium tetrafluoroborate (NOBF$_4$) to exchange the native ligands (oleyl alcohol) on the surface of γ-Fe$_2$O$_3$ NPs and charge stabilized them by BF$_4^−$ anions as shown in Figure 1 [41]. The advantage of using ligand-stripped NPs is that their surface OH$^−$ groups may provide high solubility in polar solvents while ensuring close proximity between them during drying process. A well-connected NP network is achieved using this method. The BF$_4^−$-capped γ-Fe$_2$O$_3$ NPs can be readily dispersed in various polar solvents, such as *N, N*-dimethylformamide (DMF), acetonitrile and ethanol, producing stable colloidal solutions even for several months as shown in Figure S1. Using the spin coating technique, the bare γ-Fe$_2$O$_3$ NPs can then be self-assembled into 2D arrays of closely packed NPs with conductive bridges between them and low surface roughness, attributes that are crucial for planar p-i-n perovskite solar cells (PVSCs) applications.

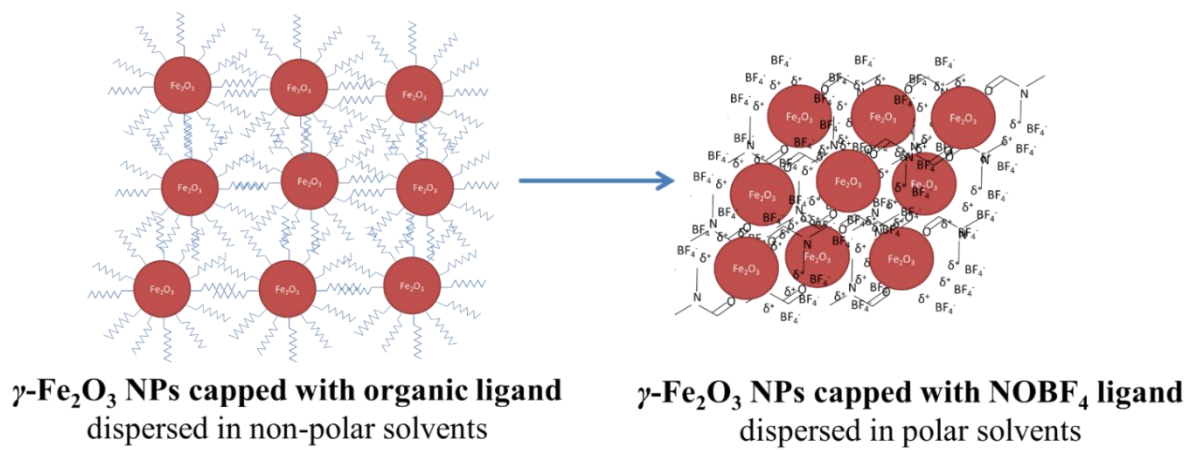

**Figure 1.** Schematic illustration of the ligand exchange method for the preparation of $BF_4^-$-capped γ-$Fe_2O_3$ NPs.

The crystal structure and morphology of the as-prepared γ-$Fe_2O_3$ NPs were characterized by X-ray diffraction (XRD) and transmission electron microscopy (TEM). The XRD pattern in Figure 2c demonstrates the formation of a cubic spinel-type structure with $P4_132$ space group symmetry (JCPDS No. 39-1356). It is worth noting that within the XRD detection limits (~3–5 wt%) no impurity crystal phases were detected (e.g., α-$Fe_2O_3$ and spinel $Fe_3O_4$ phase), indicating that the obtained NPs are single-phase γ-$Fe_2O_3$. Using the (311) peak width of the diffraction pattern and the Scherrer equation, we calculated the average crystalline size of ~5.5 nm. Figure 2a displays a typical TEM image for the γ-$Fe_2O_3$ NPs sample. The image shows spherical-shaped NPs with diameter of ~5 nm, which is consistent with the γ-$Fe_2O_3$ grain size obtained from XRD analysis. The crystal structure of the NPs was further examined by selected-area electron diffraction (SAED). Consistent with XRD studies, the SAED pattern of the γ-$Fe_2O_3$ NPs sample revealed its pure phase and high crystallinity. All the Debye-Scherrer diffraction rings can be indexed to the cubic, closely packed structure of γ-$Fe_2O_3$ (Figure 2b). To investigate the optical properties of the as-prepared γ-$Fe_2O_3$ NPs, we utilized diffuse reflectance ultraviolet-visible/near-IR (UV-vis/NIR) spectroscopy. The UV-vis/NIR spectrum shows an intense optical absorption onset in the visible region, which is associated with an optical band gap of 1.9 eV, as estimated from the Tauc plot for direct interband transitions (Figure 2d).

Thin films of γ-$Fe_2O_3$ NPs were spin-casted next on top of PC(70)BM and quartz substrates using the spin coating technique, applying the processing parameters as described in the Experimental Section. Figures S2 and S3 display the surface topography of a 15 ± 5 nm thick film of γ-$Fe_2O_3$ NPs fabricated on top of PC(70)BM and quartz substrates, respectively, as obtained by AFM line scans. On top of PC(70)BM the surface roughness is ~5 nm, which is comparable with the roughness obtained by plain PC(70)BM as well as PC(70)BM/AZO films. The γ-$Fe_2O_3$ film fabricated on quartz substrate exhibits a smooth and compact topography of only ~4.3 nm roughness. It should be noted that the development of a low roughness layer is beneficial for the photovoltaic perovskite solar cells (PVSCs) performance. Figure S4 shows the transparency of the AZO and γ-$Fe_2O_3$ layers, respectively. As shown in Figure S4, AZO layer presents high transparency in the visible region, while a slightly lower transmittance is observed for γ-Fe$_2$O$_3$ layer.

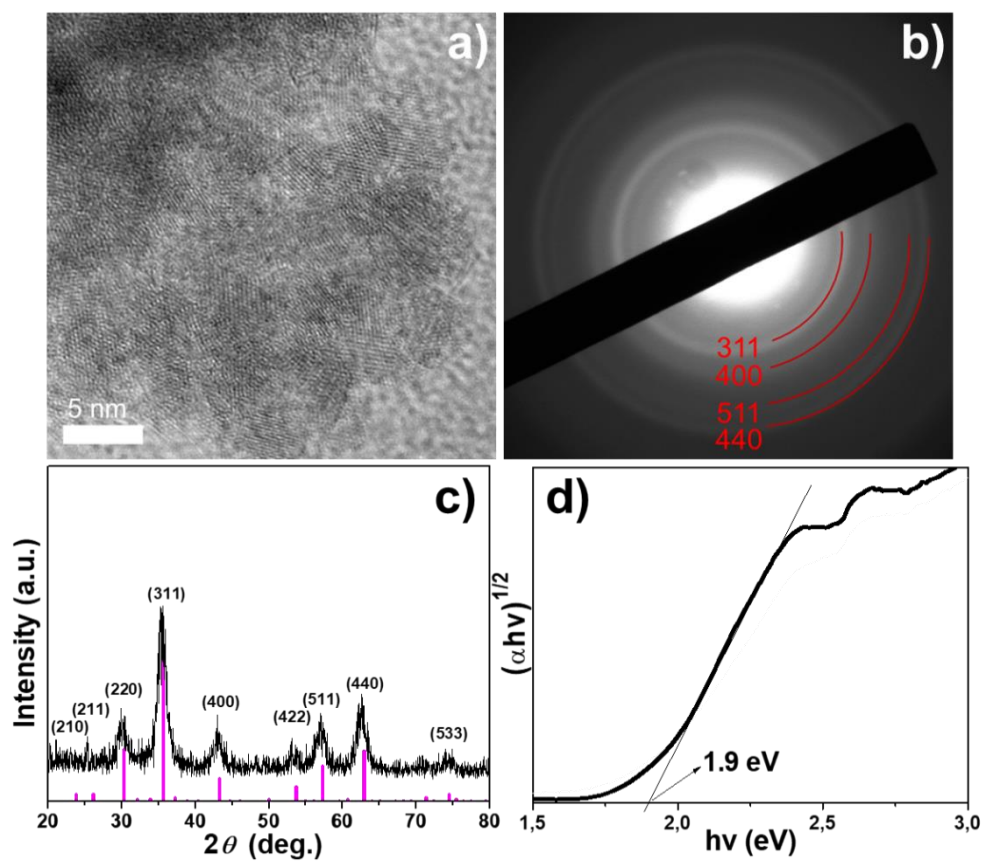

**Figure 2.** (**a**) Typical TEM image, (**b**) SAED pattern, (**c**) XRD pattern of γ-Fe$_2$O$_3$ NPs. The standard pattern of maghemite phase, γ-Fe$_2$O$_3$, (JCPDS No. 39-1346) is also illustrated for collation (magenta bars) and (**d**) UV-Vis diffuse reflectance spectra versus energy (hv) for the γ-Fe$_2$O$_3$ NPs.

Photoluminescence (PL) was performed at 550 nm excitation wavelength, to investigate the electronic band structure and charge transfer properties of γ-Fe$_2$O$_3$ NPs, as shown in Figure 3a. In the case of pure defect-free crystals, we can expect only the band edge emission in the PL spectra. However, the PL spectrum of γ-Fe$_2$O$_3$ NPs exhibits a weak peak at ~680 nm, which is a higher wavelength than band edge emission This is ascribed to defect levels in the forbidden energy gap of the nanocrystals. Electrochemical impedance spectroscopy (EIS) was then used to investigate the energy levels of γ-Fe$_2$O$_3$. The resulting Mott-Schottky plot and the corresponding fit of the linear portion of the inverse square space-charge capacitance ($1/C_{sc}^2$) as a function of potential (E) for γ-Fe$_2$O$_3$ NPs are shown in Figure 3b. It is apparent that the γ-Fe$_2$O$_3$ shows a positive linear slope, indicating n-type conductivity, where electrons are the majority carriers. Using extrapolation to $1/C_{SC}^2 = 0$, the flat-band potential ($E_{FB}$) calculated for γ-Fe$_2$O$_3$ NPs was 0.19 eV vs. NHE (pH = 7). Based on the $E_{FB}$ value and optical band gap (as obtained from UV–vis/NIR reflectance data, Figure 2d), the energy band edges for Fe$_2$O$_3$ are CB: −4.28 eV and VB: −6.18 eV vs vacuum, respectively (Figure 4b). For heavily n-type doped semiconductors such as γ-$Fe_2O_3$, it is quite reasonable to assume that the $E_{FB}$ level lies very close to the CB edge. Typically, for many n-type semiconductors, the CB edge is about 0.1–0.3 eV higher than the $E_{FB}$ potential. Thus, the position of the valence band (VB) edge was estimated from $E_{FB}$–$E_g$.

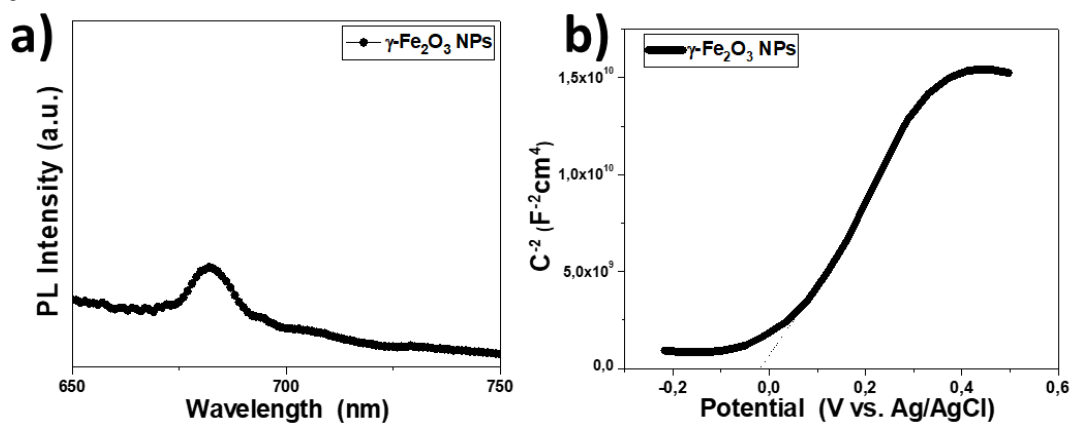

**Figure 3.** (**a**) Room-temperature PL emission spectra of the γ-$Fe_2O_3$ NPs. PL measurement was performed on γ-$Fe_2O_3$ films on the glass substrate at an excitation wavelength of 550 nm. (**b**) Mott-Schottky plot of the inverse square space-charge capacitance (1/$C_{SC}^2$) as a function of applied voltage (E) relative to the redox potential of Ag/AgCl (3 M KCl) for the γ-$Fe_2O_3$ NPs ETL.

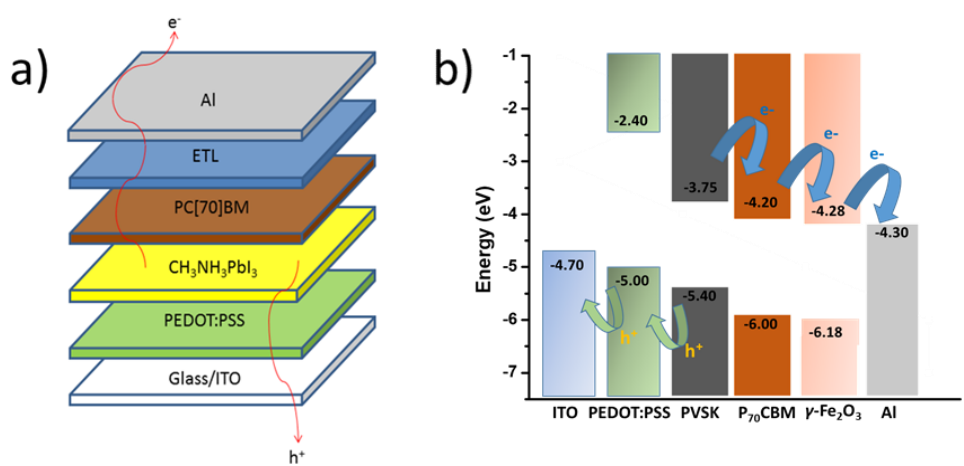

**Figure 4.** (**a**) Schematic representation of the device structure used and (**b**) the corresponding energy level diagram under study.

*3.2. Inverted Perovskite Solar Cells Heat Accelerated Lifetime Studies*

In order to test the effectiveness of γ-$Fe_2O_3$ nanoparticulate ETL for inverted structured PVSCs, completed devices were fabricated based on the general inverted device structure ITO/PEDOT:PSS (~30 nm)/$CH_3NH_3PbI_3$ (~300 nm)/ETL/Al (~100 nm), where the three different ETLs under investigation were compared: PC(70)BM (~70 nm), PC(70)BM/AZO (~40 nm) and PC(70)BM/$Fe_2O_3$ (15 ± 5 nm). A schematic representation of the device structure is shown in Figure 4a. The exact fabrication steps and materials used are described in detail in the Experimental Section. The use of γ-Fe$_2$O$_3$ in conjunction with PC(70)BM was found to have little impact on the PCE of p-i-n structured devices since the PCEs were comparable both to PC(70)BM as well as PC(70)BM/AZO. This is further highlighted in the energy level diagram shown in Figure 4b, which is based on measured values of γ-Fe$_2$O$_3$ with EIS and the literature respectively [39,42]. The representative values of PV parameters from a total of 10 devices from each structure, that described in detail above, are summarized in Table 1.

As we have previously noted, by using PC(70)BM/γ-Fe$_2$O$_3$ as electron selective contact instead of plain PC(70)BM or PC(70)BM/AZO appears to have no significant effect to the device PCE (only a slight drop in FF, from 75 to 70%, was observed). It is important to note that the focus of this work was to inspect the effect of γ-Fe$_2$O$_3$ as a solution processed interface modification for improving the stability of the inverted PVSCs. For the same reasoning we chose to use widely used and understood materials such as PEDOT: PSS as the HTL and a basic perovskite CH$_3$NH$_3$PbI$_3$ formulation as the active layer for the inverted PVSCs. The device structure and perovskite formulation used within this work offer high reliability and reproducibility, essential parameters for fabricated devices in lifetime studies. The experimental results presented were consistent on several experimental runs

**Table 1.** Photovoltaic parameters for the three different device architectures.

| Device Architecture | Voc (V) | Jsc (mA/cm$^2$) | FF (%) | PCE (%) |
|---|---|---|---|---|
| ITO/PEDOT:PSS/CH$_3$NH$_3$PbI$_3$/PC(70)BM/Al | 0.84 | 15.41 | 75 | 9.68 |
| ITO/PEDOT:PSS/CH$_3$NH$_3$PbI$_3$/PC(70)BM/γ-Fe$_2$O$_3$/Al | 0.88 | 16.53 | 69.7 | 10.13 |
| ITO/PEDOT:PSS/CH$_3$NH$_3$PbI$_3$/PC(70)BM/AZO/Al | 0.85 | 15.4 | 74.4 | 9.72 |

We have recently reported that in inverted PVSCs using PEDOT: PSS HTL, the main heat degradation pathway is the interaction of Al top metal contact with the perovskite active layer through diffusion mechanisms. We have incorporated a thick PC(70)BM (200 nm) diffusion blocking layer to improve stability. However, in some cases incorporation of thick buffer layers, has a significant negative impact on the inverted PVSCs PCE [43]. In this work, we investigate the effect of using a thin (15 ± 5 nm) γ-F$_2$O$_3$ based interfacial layer on the heat accelerated lifetime performance of inverted PVSCs. The accelerated heat lifetime performance of inverted PVSCs incorporating γ-Fe$_2$O$_3$ within the top electrode (PC(70)BM/Fe$_2$O$_3$/Al) has been compared with the more conventional PC(70)BM/Al and PC(70)BM/AZO/Al inverted PVSCs top electrodes. The conditions for the lifetime experiments were 60 °C and N$_2$ atmosphere under dark for all PVSCs presented within the paper. The normalized lifetime was plotted from the average PV parameters of 10 devices for each architecture and the results which are listed in Figure 5 were consistent on several lifetime runs for the conditions described within this work.

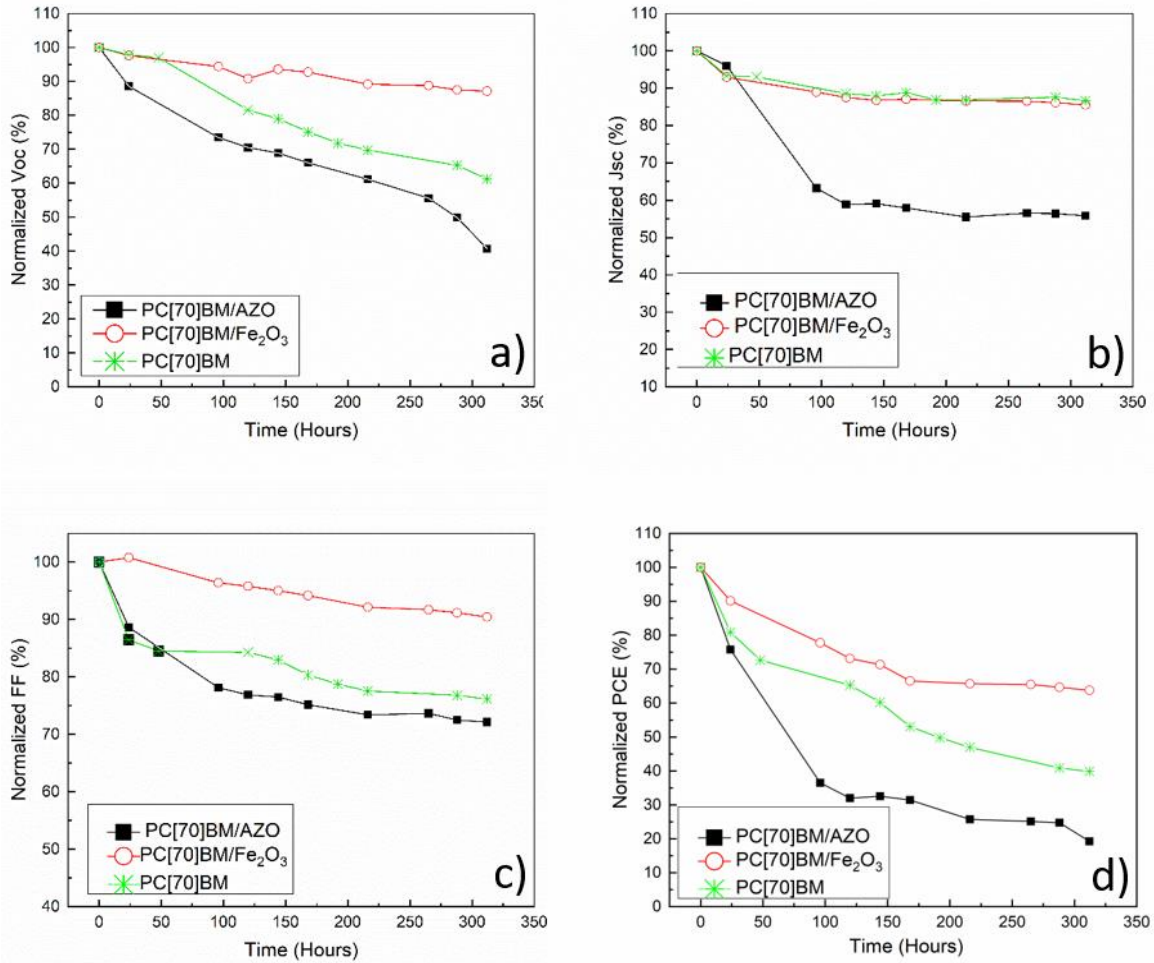

**Figure 5.** Normalized (**a**) Voc, (**b**) Jsc, (**c**) FF and (**d**) PCE for three different top electrodes: PC(70)BM/Al, PC(70)BM/AZO/Al, PC(70)BM/γ-Fe$_2$O$_3$/Al.

From the lifetime plots, there are some considerable differences between the behavior of the three different (top electrode) inverted PVSCs device architectures. The PV parameters of the PVSCs incorporating γ-Fe$_2$O$_3$ within the top electrode (PC(70)BM/γ-Fe$_2$O$_3$/Al) show considerable improvement in stability compared to devices based on PC(70)BM/Al and PC(70)BM/AZO/Al top electrode. Specifically, PC(70)BM/Fe$_2$O$_3$/Al based devices have T80 (the time the PCE reached 80% of its initial fabrication value) at 100 h of operation and has even retained 70% of their PCE after 300 h of testing, whereas devices with PC(70)BM/Al have T80 at 24 h and the PCE settled at the lowest value of 40% of their original PCE when the aging test has reached 300 h. The devices with PC(70)BM/AZO/Al have dropped to 70% at the first 24 h of aging, reaching below 40% at the 100 h mark and have settled even lower at 15% when the aging test has reached the 300 h. Figure 5 shows that Jsc in the PC(70)BM/AZO/Al-based inverted PVSCs shown the strongest drop under the heat accelerated lifetime condition presented within this paper.

*3.3. Device Performance Analysis and Degradation Mechanisms*

Complementary to the heat accelerated lifetime test, we have carried out characterization studies on representative devices for the three device architectures at the 168 h mark. We have chosen this point for characterization since it lies in the middle of the aging test and thus it would ensure that the devices showed significant differences between them while still being eligible for testing and are not completely degraded from the aging test. In particular, the devices were characterized using illuminated and dark J-V measurements as well as Mott-Schottky analysis (discussed in detail within the next section).

Figure 6a depicts the illuminated J-V characteristic for the three device architectures. The Voc change with respect to aging is very clearly observed. In particular, the device based on PC(70)BM/AZO/Al shows a very sharp drop to Voc, indicated by the shift of the J-V curve to the left. There is also a very significant sign of hysteresis in this device which could be attributed to charge traps or ion migration manifesting through the aging test [44]. Both inverted PVSCs architectures using PC(70)BM/Al and PC(70)BM/Fe$_2$O$_3$/Al show negligible signs of hysteresis. However, the devices without γ-Fe$_2$O$_3$ interface modification (PC(70)BM/Al) show a significant drop to Voc (although not as severe as in PC(70)BM/AZO/Al-based inverted PVSCs), while inverted PVSCs using PC(70)BM/Fe$_2$O$_3$/Al provide clearly more stable Voc values under heat accelerated lifetime conditions. Dark J-V plots are shown in Figure 6b. From the dark JV plots the series resistance (Rs and Shunt resistance (Rsh) were estimated (Table 2). The devices with PC(70)BM/AZO/Al top electrode show significantly reduced Rsh from 36.26 to 5.56 kΩ and increased leakage current after aging compared to the other two top electrode architectures under investigation. Furthermore, the diode characteristic is completely lost in devices with PC(70)BM/AZO/Al top electrode. In comparison, the device with PC(70)BM/Al has also shown a decrease to Rsh from 37.04 to 11.24 kΩ and increase in leakage current which is not as severe as the PC(70)BM/AZO/Al-based inverted PVSCs. Importantly, the inverted PVSCs with PC(70)BM/Fe$_2$O$_3$/Al top electrodes show almost no change after aging regarding Rsh and leakage current (a parameter which is shown the quality of the inverted PVSCs incorporating the proposed γ-Fe$_2$O$_3$ interface modification). Both devices with PC(70)BM and PC(70)BM/Fe$_2$O$_3$ ETLs seem to retain the diode characteristic after aging. This trend follows the observation of Voc drop, where devices with decreased Rsh and increased leakage current also show a proportional drop to their Voc.

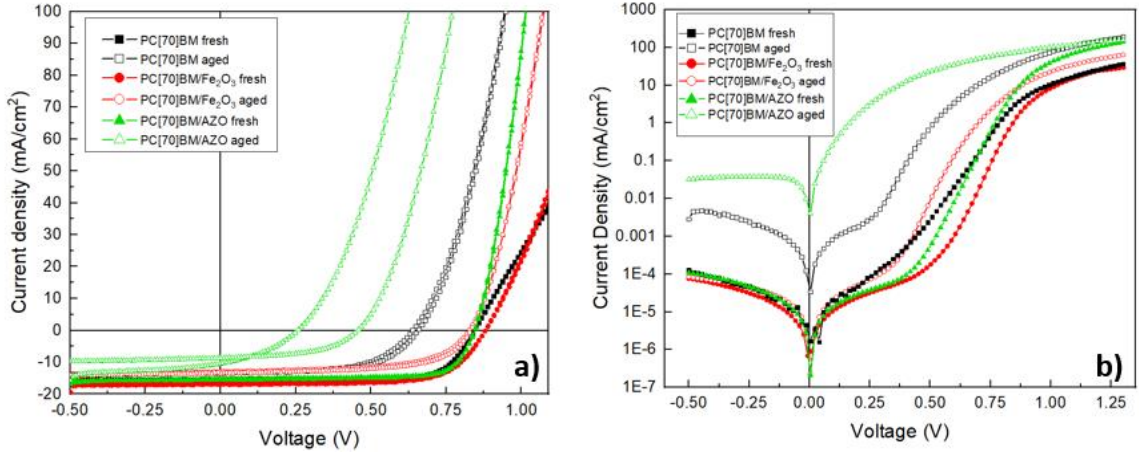

**Figure 6.** (**a**) Illuminated J-V characteristics and (**b**) dark J-V characteristics, respectively.

**Table 2.** Rs and Rsh values for fresh and aged devices.

| Device Architecture | Rs Fresh ($\Omega$) | Rsh Fresh (k$\Omega$) | Rs Aged ($\Omega$) | Rsh Aged (k$\Omega$) |
|---|---|---|---|---|
| ITO/PEDOT:PSS/CH$_3$NH$_3$PbI$_3$/PC(70)BM/Al | 9.78 | 37.04 | 2.45 | 11.24 |
| ITO/PEDOT:PSS/CH$_3$NH$_3$PbI$_3$/PC(70)BM/$\gamma$-Fe$_2$O$_3$/Al | 11.8 | 38.23 | 7.18 | 33.66 |
| ITO/PEDOT:PSS/CH$_3$NH$_3$PbI$_3$/PC(70)BM/AZO/Al | 4.83 | 36.26 | 4.7 | 5.56 |

Following the observations from our previous work based on absorption spectra and XRD measurements [43], as well as other reports in the literature [18], the MAPI perovskite active layer was not identified as the major degradation pathway for inverted perovskite solar cells under accelerated heat lifetime conditions [43]. The two main mechanisms that could be related to the degradation of the devices are (i) diffusion of halide ions to the metal electrode and (ii) diffusion of metal atoms to the CH$_3$NH$_3$PbI$_3$. Both these mechanisms were observed by Fang et al. where diffusion of I$^-$ towards the Ag electrode, as well as migration of Ag into CH$_3$NH$_3$PbI$_3$ under ambient humidity and light soaking for 200 h [14]. In particular, diffusion of I$^-$ ions are common in such devices due to the small activation energy (~0.29 eV), high concentration gradient ($10^{15}$ cm$^{-4}$) and fast diffusion coefficient ($3.1 \times 10^{-9}$ cm$^2$s$^{-1}$) [45] which is further accelerated at higher temperatures [46]. Metal ions also exhibited diffusion towards the perovskite under accelerated heat conditions. It was recently reported that upon heating at 70 °C, Au atoms can travel through spiro-MeOTAD HTL and migrate towards the perovskite layer, negatively affecting the overall device performance. [19] In previous works buffer layers such as Cr$_2$O$_3$/Cr [47] and SnO$_x$ [22] were used in conjunction with PCBM to reduce the degradation by blocking the aforementioned diffusive degradation products of inverted PVSCs under accelerated heat conditions and constant light illumination.

Another major concern that has been recently placed is the negative effect that doping can have on the stability of PSCs. Recently a comprehensive study was performed by N. Tessler and Y. Vaynzof using a model solar cell system based on electron-ion conducting perovskite active layer. In that work it was shown that undoped charge selective layers (CSLs) can be used to significantly decrease the voltage drop across the perovskite active layer and therefore suppress the ion accumulation through the device, preventing hysteresis and degradation of perovskite solar cells [48]. On contrary using doped CSLs the potential drop across the perovskite active layer was reported to be significantly larger compared to undoped CSLs and therefore the suppression of ion accumulation is expected to be less efficient [48]. Although in our case we compare two ETLs that one is doped (AZO) and another that is undoped ($Fe_2O_3$), it is worth mentioning the role of the PC(70)BM passivation/buffer layer that lies between the AZO and the perovskite active layer. Since The PC(70)BM layer is inherently undoped, ionic accumulation issues from doping related to AZO should not be dominant [48]. The fact that devices with AZO have shown the worst stability led us to search for a different explanation. The instability of AZO in p-i-n PSCs has also been previously reported by K. Brinkmann et al. [22]. It is reported in several works that zinc oxide (ZnO) can negatively impact the stability of $CH_3NH_3PbI_3$ upon annealing through the deprotonation of $MA^+$ caused by residual $OH^-$ on the ZnO nanoparticles surface that are often present. It was also reported that PCBM as a buffer layer can reduce, but not completely avoid this decomposition [21,49]. This correlates with our observation from the lifetime plots in Figure 5b, showing a strong reduction in Jsc values for devices with AZO, pointing towards the degradation of the PVSK active layer whereas both PC(70)BM and PC(70)BM/$\gamma$-$Fe_2O_3$ show negligible drop to Jsc.

A crucial factor for the stability and the quality of halide perovskite crystallization is the acidic or basic character of the metal oxides, which are used as HTLs or ETLs in the perovskite solar cells (PVSCs). The acidic or basic nature of metal oxides depends on the corresponding isoelectric point (IEP). It has been observed that these hydroxyl groups on the interfaces of metal oxides with basic nature cause decomposition of the hybrid halide perovskite through the deprotonation of the methylamine, leading to $PbI_2$. This decomposition mechanism is not observed in more acidic metal oxides (such as $TiO_2$ and ITO) [50]. In the literature, $\gamma$-$Fe_2O_3$ NPs interfaces are typically acidic (with isoelectric points in the range of 5.5−6.7) [51], whereas ZnO NPs interfaces tend to be more basic (with isoelectric points >8.7) [52]. This suggests that the deprotonation of the $MA^+$ (pKa of ~10.6) by the ZnO surface may be a possible cause of the thermal decomposition [53]. Since ZnO is the main component of AZO, the experimental results presented within this paper indicate that in contrast to the basic nature of aluminum-doped ZnO (AZO), the more acidic surface of $\gamma$-$Fe_2O_3$ assist on preventing negative chemical interactions with the perovskite active layer resulting to inverted perovskite solar cells with improved Jsc stability under heat accelerated lifetime studies. Furthermore, by the incorporation of $\gamma$-$Fe_2O_3$ interfacial layer on top of PC(70)BM any surface defects of inconsistent film coverage from the PC(70)BM film can be passivated by the γ-Fe$_2$O$_3$ interfacial layer improving the stability of the inverted perovskite photovoltaics devices under accelerated heat lifetime conditions.

*3.4. Impedance Spectroscopy Device Characterization and the Effect of Interfacial Trap States*

Using Mott-Schottky analysis we have tried to understand why the incorporation of γ-Fe$_2$O$_3$ interfacial layer within the top electrode (PC(70)BM/γ-Fe$_2$O$_3$/Al) has resulted in better stability compared to more conventional top electrodes such as PC(70)BM/Al and PC(70)BM/AZO/Al. Capacitance voltage (CV) techniques are useful characterization tools that can help distinguish processes that manifest in the active layer from the ones in the contacts. At reverse bias full depletion occurs and the capacitance corresponds to the geometric capacitance (Cg) determined by the dielectric constant of the perovskite [47]. The point of intersection with the x-axis denotes the extracted built-in voltage (E$_{Vbi}$) value which depends on the energy equilibration at the contacts. Mott Schottky plots are based on the depletion approximation, where we consider that the capacitance in the space charge region is purely due to doping and there are no free charges. It is interesting to note that in general this approximation doesn't hold true close to the metal contacts where charges tend to accumulate, thus the value of EVbi calculated from the Mott-Schottky analysis is not the true value of Vbi, but can still be used for comparative studies [47]. Mott-Schottky analysis has been previously utilized for the quantification of charge traps in PSCs by studying the response of charged carriers to the AC signal that is applied in the CV measurement. When using relatively high frequencies (HF) at ~100 kHz only free charges at the conduction band (Ec) can follow the AC voltage perturbation applied in CV measurements and can contribute to the E$_{Vbi}$ value. When the frequency is dropped to low frequencies (LF) at ~500 Hz we are effectively approaching DC conditions, where de-trapping of charges can occur and therefore contribute to E$_{Vbi}$ and as a result of Voc [54]. Therefore, the presence of trapped charges can contribute to the overall E$_{Vbi}$ value, effectively altering it, depending on the frequency used for the Mott-Schottky analysis [55]. Using this information we have performed a comparative study where the E$_{Vbi}$ change between HF and LF was correlated with the population of traps inside freshly prepared devices.

From Figure 7a and 7b we have used linear interpolation to extract the value of E$_{Vbi}$ from the point of intercept of the linear part in the Mott-Schottky plot with the x-axis. Some clear differences are observed in the values of E$_{Vbi}$ shown in Table 3. First by comparing the E$_{Vbi}$ drop of fresh devices incorporating PC(70)BM/Al with PC(70)BM/γ-Fe$_2$O$_3$/Al and PC(70)BM/AZO/Al we see that they exhibit the largest drop at 0.73 V compared to 0.40 V and 0.30 V for the PC(70)BM/γ-Fe$_2$O$_3$/Al and PC(70)BM/AZO/Al, respectively. This indicated that there is a larger population of charge traps in fresh devices with PC(70)BM/Al compared to PC(70)BM/γ-Fe$_2$O$_3$/Al and PC(70)BM/AZO/Al.

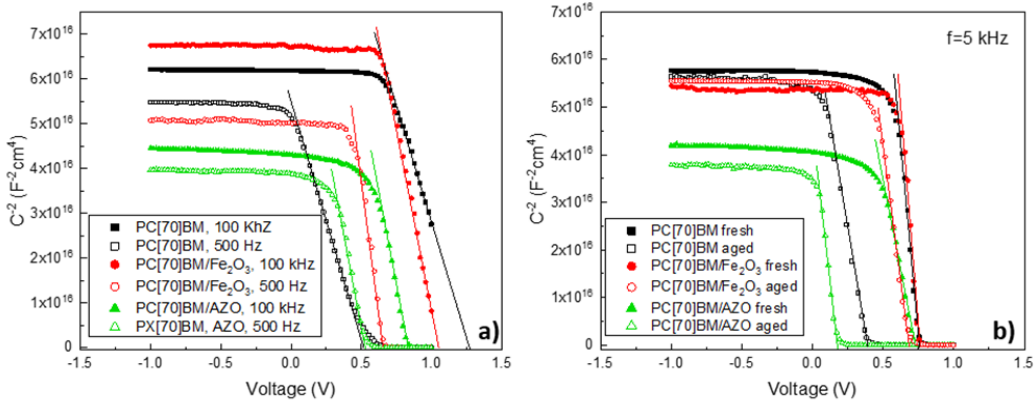

**Figure 7.** (**a**) Mott Schottky measurements of fresh devices at f = 100 kHz and f = 500 Hz and (**b**) fresh and aged devices at f = 5 kHz.

Identifying the exact nature of charge traps would require further complex characterization techniques and therefore was beyond the scope of this paper, but since the only change in the device structure is within the top electrode, the traps are present mainly in the PC(70)BM/Al interface. Charge traps were reported to negatively impact the stability of PSCs from the induction of the local electric field [56]. We propose in this paper that by incorporating a suitable γ-$Fe_2O_3$ interfacial layer more intimate interface with the top metal contact can be achieved.

**Table 3.** Extracted built-in voltage $E_{Vbi}$ voltage drop between high frequencies (HF)- and low frequencies (LF) of fresh devices and drop between fresh (as prepared) and aged (degraded) devices.

| Device Architecture | $E_{(Vbi)}$ Drop between HF and LF of Fresh Devices (V) | $E_{(Vbi)}$ Drop between Fresh and Aged Devices (V) |
|---|---|---|
| ITO/PEDOT:PSS/$CH_3NH_3PbI_3$/PC(70)BM/Al | 0.73 | 0.39 |
| ITO/PEDOT:PSS/$CH_3NH_3PbI_3$/PC(70)BM/γ-$Fe_2O_3$/Al | 0.4 | 0.09 |
| ITO/PEDOT:PSS/$CH_3NH_3PbI_3$/PC(70)BM/AZO/Al | 0.3 | 0.53 |

In order to test the $E_{Vbi}$ drop between fresh and aged devices, we have kept the frequency constant at f = 5 kHz. When choosing an appropriate frequency for Mott-Schottky analysis it should usually coincide in the high-frequency plateau of the relevant capacitance frequency (CF) plot [57]. We also note that at 5 kHz the $E_{Vbi}$ value for all three device architectures under investigation is similar, which is in agreement with their similar measured $V_{oc}$ values. CF measurements were performed using this particular perovskite formulation, fabricating devices with the general structure ITO/PEDOT:PSS/$CH_3NH_3PbI_3$/PC(70)BM/AZO/Al in our previous work and therefore 5 kHz is an appropriate choice for our particular measurements [58]. After aging, the trend in $E_{Vbi}$ drop at 5 kHz correlates well with the $V_{oc}$ drop that we see in the lifetime plots from Figure 5a, where the smallest E$_{vbi}$ drop is observed by devices incorporating PC(70)BM/γ-Fe$_2$O$_3$/Al at 0.09 V followed by PC(70)BM/Al at 0.39 V, whereas the PC(70)BM/AZO/Al devices exhibit the largest E$_{vbi}$ drop at 0.53 V. As discussed in detail in the above section AZO is not a suitable material for interface modification due to interaction with CH$_3$NH$_3$PbI$_3$ negative affecting the corresponding inverted PVSK PVs lifetime performance. By using the proposed γ-Fe$_2$O$_3$ interface modification between the ETL and top metal contact effective interfaces for the inverted perovskite top electrode (PC(70)BM/γ-Fe$_2$O$_3$/Al) can be achieved.

## 4. Conclusions

In conclusion, we have shown that colloidal synthesized γ-Fe$_2$O$_3$ can be effectively used to provide solution processed nanoparticulate interfacial layer for the top electrode of inverted PVSCs. Inverted PVSCs with PC(70)BM/Fe$_2$O$_3$/Al top electrode yield comparable PCE value to the more conventional inverted PVSCs top electrodes such as PC(70)BM/Al and PC(70)BM/AZO/Al. The proposed inverted PVSCs γ-Fe$_2$O$_3$ top electrode interface modification (PC(70)BM/Fe$_2$O$_3$/Al) results in improved stability under accelerated heat conditions at 60 °C and N$_2$ atmosphere. On the contrary PC(70)BM/AZO/Al device has shown the worst stability most likely due to the deprotonation of MA$^+$ induced by the basic nature of AZO compared to the more acidic γ-Fe$_2$O$_3$ interface modification (PC(70)BM/γ-Fe$_2$O$_3$/Al). Compared to plain PC(70)BM/Al we have shown that PC(70)BM/γ-Fe$_2$O$_3$/Al could assist in better coverage and passivation of surface defects related to top electrode interfaces. Overall, PC(70)BM/γ-Fe$_2$O$_3$/Al results in less population of charge traps that mainly lie in the PC(70)BM/Al interface and, therefore, γ-Fe$_2$O$_3$ interface modification provides a better top electrode for inverted PVSCs. From impedance measurements we have shown that incorporating γ-Fe$_2$O$_3$ provides a negligible loss in Evbi, as shown in the Mott-Schottky plot, which follows the reduced Voc loss observed within the reported heat accelerated lifetime performance. The results presented highlight the importance of intimate interfaces for the development of PVSCs with long term stability.

**Acknowledgments:** This research was funded by European Research Council (ERC) under the European Union's Horizon 2020 research and innovation program (grant agreement No 647311).

# Supporting Information

## Nanoparticulate Metal Oxide Top Electrode Interface Modification Improves the Thermal Stability of Inverted Perovskite Photovoltaics

By Ioannis T. Papadas,[1] Fedros Galatopoulos,[1] Gerasimos S. Armatas,[2] Nir Tessler,[3] and Stelios A. Choulis*,[1]

[1] Molecular Electronics and Photonics Research Unit, Department of Mechanical Engineering and Materials Science and Engineering, Cyprus University of Technology, Limassol, 3603 (Cyprus).
[2] Department of Materials Science and Technology, University of Crete, Heraklion 71003, Greece
[3] Sara and Moshe Zisapel Nano-Electronic Center, Department of Electrical Engineering, Technion-Israel Institute of Technology, Haifa 32000, Israel

*Corresponding Author: Prof. Stelios A. Choulis

E-mail: stelios.choulis@cut.ac.cy

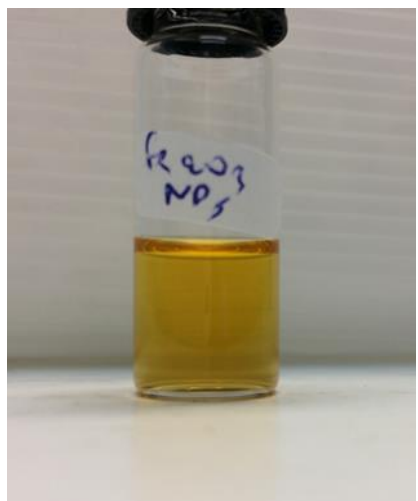

**Figure S1:** Dispersion of γ-$Fe_2O_3$ NPs in ethanol.

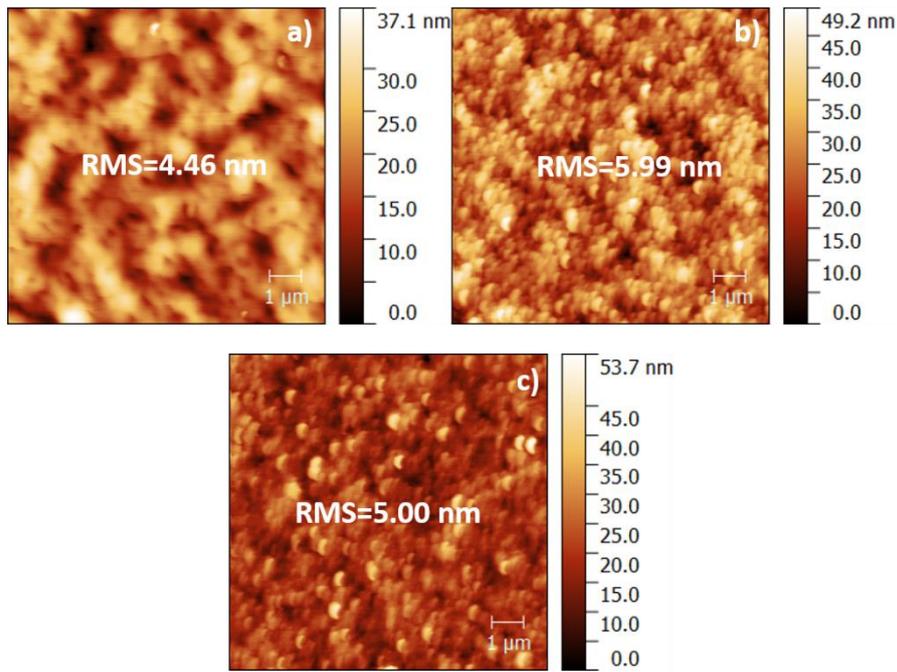

**Figure S2:** AFM data in 10x10 μm magnification for **a)** ITO/PEDOT:PSS/ $CH_3NH_3PbI_3$/PC[70]BM, **b)** ITO/PEDOT:PSS/ $CH_3NH_3PbI_3$/PC[70]BM/AZO and **c)** ITO/PEDOT:PSS/ $CH_3NH_3PbI_3$/PC[70]BM/γ-$Fe_2O_3$ films, respectively.

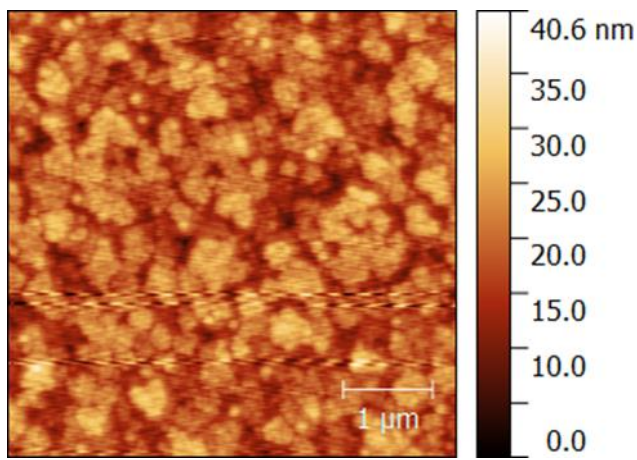

**Figure S3:** AMF data in 5x5 μm magnification for γ-$Fe_2O_3$ film on quartz substrate.

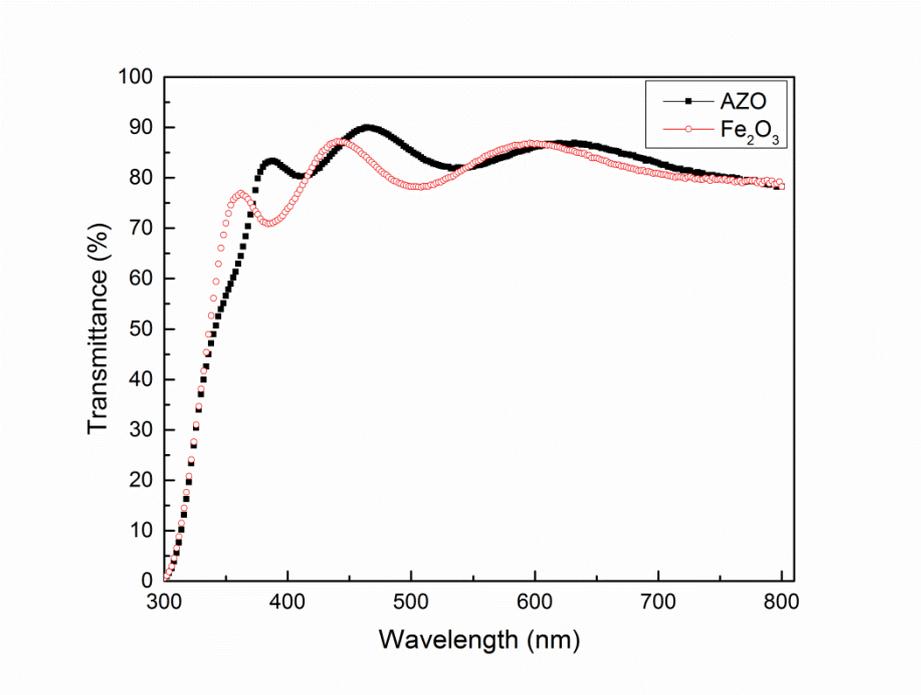

**Figure S4:** Transmittance spectra of AZO and γ-$Fe_2O_3$ films on quartz substrates